# Implementation of ICH E9 (R1): a few points learned during the COVID-19 pandemic

Yongming Qu*, Ilya Lipkovich

Eli Lilly and Company, Indianapolis, IN 46285, USA

*Correspondence: Yongming Qu, Department of Biometrics, Eli Lilly and Company, Lilly Corporate Center, Indianapolis, IN 46285, USA. Email: qu_yongming@lilly.com

Abstract

The current COVID-19 pandemic poses numerous challenges for ongoing clinical trials and provides a stress-testing environment for the existing principles and practice of estimands in clinical trials. The pandemic may increase the rate of intercurrent events (ICEs) and missing values, spurring a great deal of discussion on amending protocols and statistical analysis plans to address these issues. In this article we revisit recent research on estimands and handling of missing values, especially the ICH E9 (R1) on Estimands and Sensitivity Analysis in Clinical Trials. Based on an in-depth discussion of the strategies for handling ICEs using a causal inference framework, we suggest some improvements in applying the estimand and estimation framework in ICH E9 (R1). Specifically, we discuss a mix of strategies allowing us to handle ICEs differentially based on reasons for ICEs. We also suggest ICEs should be handled primarily by hypothetical strategies and provide examples of different hypothetical strategies for different types of ICEs as well as a road map for estimation and sensitivity analyses. We conclude that the proposed framework helps streamline translating clinical objectives into targets of statistical inference and automatically resolves many issues with defining estimands and choosing estimation procedures arising from events such as the pandemic.

The COVID-19 pandemic started in early 2020 has had a great impact on ongoing clinical trials from multiple aspects, including study treatment interruptions, treatment or study discontinuations, and missing clinical visits due to COVID-19 control measures (patient or site personnel's quarantine and travel restrictions, site closure, disruption on drug supply chain, or transportation) and COVID-19 illness. This has spurred many discussions within and across pharma companies, often resulting in amending the estimands and estimation methods in the protocol and/or statistical analysis plan for ongoing studies [1,2]. The pandemic provides an opportunity for stress-testing of the estimand framework and its implementation. With that goal in mind we identified several gaps in the interpretation or implementation of ICE E9 (R1) [3]. Here, we have summarized our findings, often in a polemical manner, which may help researchers better understand and implement the estimand framework within the clinical community. For this purpose, we gave each subsection a title reflecting specific perspective.

**1. Lack of generalizability in "treatment policy" strategy**

The potential outcome (PO) framework [4,5] was used by Lipkovich et al. [6] to define causal estimands. Let $Y_i$ denote the outcome of interest and $Y_i(a, b)$ denote the PO with assigned treatment regimen $a$, but actually taking treatment regimen $b$ during the study. Assume we have only two treatment regimens of interest in a study and let $A_i$ denote the treatment regimen to be studied, such that $A_i = 0$ for the control treatment and $A_i = 1$ for the experimental treatment. Let $A_i^* = \{A_i, g_i(Z_i(A_i))\}$ be the treatment regimen (policy) patient $i$ takes (which generally is not precisely defined in the protocol), where $Z_i$ (possibly multidimensional) is postbaseline intermediate outcomes that affect treatment changes captured by patient-specific function $g_i(\cdot)$. Using the language of PO, the estimand defined by his treatment policy strategy is an average treatment effect (ATE) of

$$E\left\{Y_i\left(1, g_i\big(Z_i(1)\big)\right) - Y_i\left(0, g_i\big(Z_i(0)\big)\right)\right\}.$$

Treatment policy strategy essentially compares outcomes (across randomized groups) associated with actual treatment regimen (or "policy") not only driven by individual patients' outcomes but also having patient-specific rather than common rules for changing treatment $g_i(\cdot)$. The treatment policy strategy is different from the *dynamic treatment regimens* (DTR) [7,8] also referred within the causal community as "treatment policies" in which the time-varying treatment regimens are defined based on evolving patients' outcomes using a rule (regimen) $g(\cdot)$ (without the subscript *i*) that is *common* across all patients. Therefore, the estimand for DTR is defined as:

$$E\left\{Y_i\left(1, g\big(Z_i(1)\big)\right) - Y_i\left(0, g\big(Z_i(0)\big)\right)\right\}.$$

We emphasize the difference between the DTR and the treatment policy estimand because it is not well understood within the clinical trialist community.

One argument for using the treatment policy strategy is that it may apply to real clinical practice. However, except for pragmatic studies [9], the difference in the settings (e.g. frequency of visit, diligence of follow-up, allowed concomitant medications, etc.) between clinical studies and the real-world setting is generally large. A clear description of treatment regimens (including the treatment of interest and specific requirements of compliance) will allow clinicians to better understand the difference between treatment regimens implemented in clinical trials and treatment regimens observed in real-world practice. This will also make it easier to relate estimates of treatment effect from clinical trials to specific clinical settings. Treatment policy strategy, which takes the observed outcomes ignoring ICEs that are not clinically defined as part of the treatment regimen, will make such interpretation difficult. The COVID-19 pandemic leading to increased ICEs further exacerbates the gap between treatment policy estimand and a real clinical practice.

## 2. A strong case for using hypothetical strategies

ICH E9 (R1) mentions hypothetical strategies as one generic type of strategy for handling ICEs, indicating in plain English several scenarios when they may be useful (e.g. when interest is in estimating what would have happened if the ICE of interest did not occur contrary to the fact) but does not precisely define different hypothetical strategies using mathematical language. As a supplement to the guidance document, we discuss three types of hypothetical strategies. The first causal treatment difference for a subset of patients (*S*), if they would have adhered to their assigned treatments, is the ATE in the response between the two potential treatments averaged across all patients in *S*:

$$N_S^{-1} \sum_{i=1}^{N_s} E[Y_i(1,1) - Y_i(0,0) \,|S],$$

where $N_s$ is the sample size for $S$. In causal literature, this type of causal effect is often called a *controlled direct effect* of treatment where "controlled" means we force the ICE *not* to occur and the initial treatment to continue [10]. We call this hypothetical strategy the *controlled direct hypothetical* (CDH) strategy. With this strategy, the estimand is the treatment difference if patients *would have* adhered to the designed treatment regimen. This strategy is most applicable when patients discontinue treatment for reasons unrelated to the experimental treatment, e.g., due to administrative reasons including the COVID-19 control measures. The CDH strategy may also be applicable for the ICEs related to using rescue medications due to ethical reasons and using concomitant medications to treat COVID-19 illness, which could potentially impact the outcome.

In these cases, one may be interested in the PO if the patient would not have used the (rescue) concomitant medications, as the (rescue) concomitant medications in the clinical trials may not reflect the real world or "normal circumstances." The CDH strategy may also be applied to prolonged treatment interruption or treatment discontinuation due to COVID-19 illness. As the COVID-19 illness does not occur under normal circumstances, one is naturally interested in estimating POs and associated treatment effect in the absence of the COVID-19 pandemic.

The second type of hypothetical strategy is interested in the POs assuming patients who experience ICEs (e.g., ICEs due to an adverse event [AE]) would have no benefit: as if the patients were left untreated starting from randomization. In this case, the ATE estimand can be written as

$$E\left[\left\{Y_i(1,-1)\Delta_i(1)+Y_i(1,1)\left(1-\Delta_i(1)\right)\right\}-\left\{Y_i(0,-1)\Delta_i(0)+Y_i(0,0)\left(1-\Delta_i(0)\right)\right\}\right],$$

where "$-1$" in the second argument of $Y_i(\cdot,\cdot)$ indicates no treatment received and $\Delta_i(a)$ is the ICE indicator (0 for no ICE and 1 for ICE occurring). We call this hypothetical strategy the *no treatment hypothetical (NTH) strategy*.

The third type of hypothetical strategy targets POs if the patient takes the medication until the ICE and then stops taking the medication. Using the PO language, the ATE estimand is defined as:

$$E\left[\left\{Y_i(1,g_i(T_i(1)))\Delta_i(1)+Y_i(1,1)\left(1-\Delta_i(1)\right)\right\}-\left\{Y_i(0,g_i(T_i(0)))\Delta_i(0)+Y_i(0,0)\left(1-\Delta_i(0)\right)\right\}\right],$$

where $T_i(a)$ is the time to the ICE under treatment $a$ and $g_i\left(T_i(\mathrm{a})\right)$ is the treatment regimen: taking treatment $a$ until the occurrence of the ICE and then having no access to treatment until a specified assessment time. This strategy assumes patients may still benefit from or be harmed by the treatment even though they discontinue the treatment earlier. We call this strategy the *partial treatment hypothetical* (PTH) strategy.

The NTH and PTH strategies may be appropriate for handling discontinuations of the study treatment due to tolerability or other AEs occurring under normal circumstances, where such patients are assumed to have no or partial benefits from the treatment.

## 3. Avoiding composite strategies by explicitly specifying all the components of a composite endpoint

The composite strategy may have been most popular in areas where the endpoint of interest is binary (failure or success), so an ICE can be easily "defined away" as a "failure" or non-response. Since in the composite strategy certain ICEs are incorporated into the outcome, it is more appropriate to explicitly include these events as components of the composite endpoint instead of classifying these events as ICEs.

Let us illustrate this using an example from clinical trials in rheumatoid arthritis. The primary endpoint is ACR20 at Week 12 [11], where ACR20 is an indicator of 20% improvement in the scale of American College of Rheumatology (ACR). A composite endpoint that treats this ICE of death or use of rescue medication as treatment failure is generally used, as described in Burmester et al. [11]. However, such a composite does not reflect the primary endpoint of ACR20 (which has nothing to do with the use of rescue medication). If it is reasonable to consider the use of rescue medication as an indication of treatment failure, one can revise the objective and the endpoint to read as "the proportion of patients with treatment success indicated by achieving the ACR20 goal at 12 weeks without use of rescue medication before 12 weeks," which would be, in our opinion, a clinically meaningful endpoint. Therefore, by explicitly incorporating all the components in the definition of an endpoint, we do not need a special "composite strategy".

**4. Principal stratum (PS) strategies should be used for defining sub-populations, not for handling ICEs**

A PS is a subset of the population defined by a PO for a post-randomization variable. As an example, suppose we are interested in estimating the treatment effect for patients with the postbaseline biomarker $S > c$ when treated with $A = 1$ throughout the study. Then, the PS is defined using the PO language $\{i\colon S_i(1,1) > c\}$ and the estimand for this PS if all patients in this stratum would adhere to the experimental treatment (using the CDH strategy) is

$$E\{Y_i(1,1) - Y_i(0,0) | S_i(1,1) > c\}.$$

Note the PS population can theoretically be combined with any strategy for handling ICEs. For example, we can define an estimand for this PS using a treatment policy strategy based on the notation in Section 2:

$$E\left\{Y_i\left(1, g_i\big(Z_i(1)\big)\right) - Y_i\left(0, g_i\big(Z_i(0)\big)\right)\middle| S_i(1,1) > c\right\}.$$

Therefore, PS is a more suitable strategy to define a (hypothetical) sub-population, rather than a strategy to handle ICEs (see a similar argument in Reference [12]). The reason one may think of using a principal stratum as a special strategy for handling ICEs is probably because a PS is often defined by an (intercurrent) event, e.g. CACE (complier average causal effect) for those who can be compliant to both treatments [13,14, SACE (survivor average causal effect) for patients who can survive under both treatments [15], and the adherer average causal effect for patients who can adhere to one or both treatments without ICEs [17]. Importantly, defining the target population in terms of a principal stratum based on an ICE (e.g. treatment discontinuation) does not provide by itself a strategy for handling this ICE, which requires defining potential outcomes of treatment regimen(s) of interest as well as handling possible missing values caused by the ICE.

**5. A mix of strategies for handling ICEs and resulting missing values should be used more broadly**

ICH E9 (R1) clearly states that the choice of strategy to handle an ICE should be based on the cause of ICEs. However, it does not explicitly advocate *using a mix of strategies* in handling ICEs within the same study. The use of more than one strategy in handling ICEs becomes more common [17, 18]. The assumption for handling the corresponding missing values should also be based on the causes and strategies of handling ICEs, or on the reason for missingness. The causes of the ICEs should be collected as accurately as possible during the clinical trials.

As an example, Figure 1 provides an illustration of using a mix of strategies in handling ICEs and missing values. This figure does not cover the situation when a single ICE may be caused by multiple reasons [19]. For example, one patient who feels the efficacy is not improved as expected while experiencing a mild AE may choose to discontinue the study medication. In this case, the cause of the treatment discontinuation is both efficacy and safety. One may define the priority order of ICEs (e.g. discontinuation due to safety as having a higher order than discontinuation due to efficacy) and use the cause with the highest order to determine the strategy for handling the ICE in question and resulting missing data (if applicable) (e.g., see Reference [18]).

## Summary and conclusions

The COVID-19 pandemic spurs discussion in the clinical trial community regarding revising protocols and statistical analysis plans to address the ICEs and missing values related to the pandemic. The very fact that so many studies impacted by the pandemic require amending their protocols and SAPs is a sign that the framework may need further improvement and clarification. In this article we provided a few discussion points for clarification and issues in interpretation or implementation in the estimand and estimation framework in ICE E9 (R1). We summarize the key points in the following:

- The principal stratum is a method to define the population for the estimand, not for handling ICEs, therefore defining a PS-based population should be followed by specifying a strategy for handling ICE's and resulting missing data. .
- The treatment(s) of interest needs to be defined explicitly prior to defining ICEs. Events that are part of treatment(s) of interest should not be classified as ICEs.
- The ICEs should be predominately handled by hypothetical strategies including (but not limited to) three hypothetical strategies: CDH, NTH and PTH strategies (Section 3).
- If a post-randomization event of interest is part of an endpoint, it should be explicitly specified as one component of the composite endpoint rather than considered an ICE. Therefore, the composite strategies should generally be avoided.
- Treatment policy strategy should be used primarily for studies that are pragmatic in nature.
- Each ICE should be handled with the most appropriate strategy for this type, which may naturally result in using a mix of strategies in handling ICEs within one study. It also requires diligent collection of the reasons for ICEs during the clinical study.

- To avoid confusion, strategies for handing ICE’s and resulting estimands should be defined in protocols and SAP’s using a formal causal language (such as based on potential outcomes) rather than plain English.

In conclusion, we hope the discussion points in this article can help streamline the process of choosing estimands and handling missing values in protocols and statistical analysis plans.

**Acknowledgements**

We are grateful to Stephen Ruberg and Bohdana Ratitch for numerous discussions on estimands which motivated us to write this article. We would like to thank Yu Du, Linda Shurzinske, and Dana Schamberger for a careful review of this article.

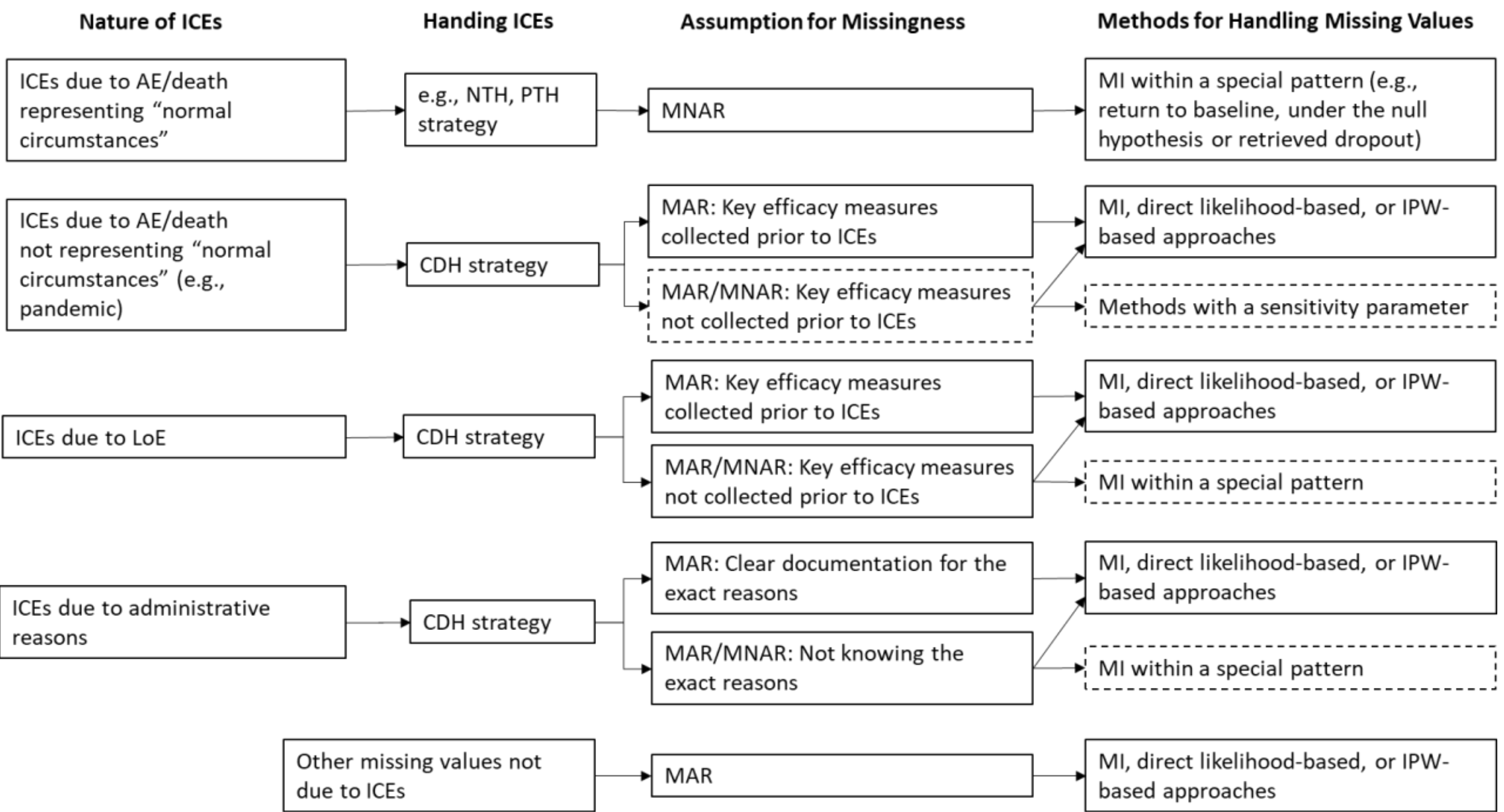

Figure 1. Handling missing values based on the nature of ICEs. ICEs that are part of treatment regimens are not included in this diagram. Missingness (or missing values) includes missing data as a result of handling ICEs by a hypothetical strategy and missing measurements of the outcome. The solid boxes are used for the primary strategy and dashed boxes are used for alternative strategies or sensitivity analyses. Abbreviations: AE, adverse events; CDH, controlled direct hypothetical; ICEs, intercurrent events; IPW, inverse probability weighting; LoE, lack of efficacy; MAR, missing at random; MI, multiple imputation, MNAR, missing not at random; NTH, no treatment hypothetical; PTH, partial treatment hypothetical.